\documentstyle[aps]{revtex}

\begin{document}

\title {On Synchronization, Persistence and Seasonality in some Spatially
Inhomogeneous Models in Epidemics and Ecology}

\author{E. Ahmed$^{1,2}$, A. S. Hegazi$^{1}$, A. S. Elgazzar$^{3,4}$ and 
H. M. Yehia$^{1}$}

\address{$^{1}$Mathematics Department, Faculty of Science, Mansoura 35516, Egypt\\
$^{2}$Mathematics Department, Faculty of Science, Al-Ain P. O. Box 17551, UAE\\
$^{3}$Max Planck Institut f\"ur Physik Komplexer Systeme, N\"othnitzer
Str. 38, D-01187 Dresden Germany\\
E-mail: elgazzar@mpipks-dresden.mpg.de\\
$^{4}$Mathematics Department, Faculty of Education, El-Arish 45111, Egypt}

\date{\today}
\maketitle

\begin{abstract}
Recent studies in ecology and epidemiology indicate that it is
important to include spatial heterogeneity, synchronization and seasonality
in the theoretical models. In this work, spatial heterogeneity is introduced via coupled map
lattices (CML) and partial differential equations. Stability and persistence of
some realistic CML are discussed. Chaos control, synchronization and
persistence are studied for some CML. Some applications in population biology
and ecology are given. A simple method for finding the sufficient
conditions for the existence of periodic solutions for differential equations
with periodic coefficients is given. This will simplify the study of
seasonality.
\end{abstract}

\section{Introduction}

Recently it has been argued \cite {ea,ll} that
synchronizing local epidemics increases the probability of its global fade
out or eradication through careful vaccination program. Similarly synchrony
between subpopulations may be an important element in species extinction.
Therefore studying spatial synchrony, chaos control to achieve synchrony,
and persistence in spatial systems is an important task. Also it was shown \cite
{ea,ll} that seasonality is an important factor that should be included. This work is
an attempt to model some of these factors simultaneously.
 
Chaos is an interesting topic with applications in many fields \cite {ho,ra}.
Most interests are to study the time evolution of systems, while in many
realistic systems spatial effects should be included. Hence both cellular automata
(CA) \cite {bo} and spatiotemporal chaos (STC) has arisen \cite {ka,cr}. In some
cases they generalize differential equations. Typically STC arises in systems with large sizes and in systems consisting
of many different components. These two conditions are satisfied in many
biological and economic systems, hence studying STC in these systems is important.

The concept of coupled map lattices (CML) was introduced by Kaneko \cite {ka} as a simple
model with the essential features of STC. It can also be used to approximate
partial differential equations. Hence it will be used here to model some
biological and economic systems. In other cases partial differential
equations will be used directly to model spatial heterogeneity.

The paper is organized as follows: In section 2, some applicable CML models are
given. A CML for a monopoly of a company producing different products is
studied. Also we have studied some problems in population biology like
prey-predator, migrating salmon in a river and Schistosomiasis. Chaos control
and chaos synchronization in CML are discussed in section 3. In section 4,
persistence in CML is studied. Section 5 is devoted for developing a method
to find the sufficient conditions for the existence of periodic solutions for
differential equations with periodic coefficients. Some conclusions are given in
section 6.

\section{Some applicable CML}

In this section, some applicable CML are presented, then chaos control and synchronization for CML
are studied. We begin by a model which is relevant to both economy and
biology. Consider a monopoly model where a company produces quantity $q(t)$ of
a goods, then the profit function is given by
\begin{equation}
\Pi (t)=q(t)[a-bq(t)-c],
\end{equation}
where $a,\;b$ are some constants and $c$ is the cost per unit of the product. The quantity
produced in the next time step is expected to be in proportion to the profit i.e.
$q(t+1)=\alpha \; \Pi (t),\;\alpha$ is a proportionality constant. Rescaling
$q(t)$, one gets the following dynamical system
\begin{equation}
\tilde{q}(t+1)=r\tilde{q}(t)[1-\tilde{q}(t)]
\end{equation}
where
\[
\tilde{q}(t)=\frac {b}{a-c}q(t), \;\; r=\alpha (a-c).
\]

Now assume that the company produces $n$ different products $\tilde{q}_{i},\;i=1,2,...,n$
(or it has different production sites), then the system (2)
is replaced by the following CML
\begin{equation}
\begin{array}{c}
q(i,t+1)=\Pi (i,t)+D \left[ \Pi (i+1,t)+\Pi (i-1,t)-2\Pi (i,t)\right],\\
\Pi (i,t)=r(i)q(i,t) \left[ 1-q(i,t) \right],
\end{array}
\end{equation}
where the tilde has been dropped for simplicity and $D$ is a constant representing
diffusion of the profits between different components of the company. Here $D$ is
considered to be small. It is straightforward to find the steady states of Eq. (3)
\begin{equation}
q(i)\simeq
q_{i}^{0}+Dq_{i}^{1},\;\;\;q_{i}^{0}=1-\frac{1}{r(i)},\;\;
\;q_{i}^{1}=\frac{q_{i-1}^{0}+q_{i+1}^{0}-2q_{i}^{0}}{r(i)-1}. 
\end{equation}

To determine the chaotic behavior of the system, local Lyapunov exponents $\lambda
_{i},\;i=1,2,...,n$ should be calculated. It is important to recall that the system
exhibits STC if $D_{L}$ is proportional to the volume of the system, where
\begin{equation}
D_{L}=K+\frac{1}{\left| \lambda _{K+1}\right| }\sum_{i=1}^{K}\lambda
_{i},\;\sum_{i=1}^{K}\lambda _{i}>0\;{\rm and}\;\sum_{i=1}^{K+1}\lambda _{i}<0. 
\end{equation}
In 1-dimensional systems this is equivalent to the condition that the number
of positive Lyapunov exponents is proportional to the size of the chain. For
$D\ll 1$, local Lyapunov exponents are given by
\begin{equation}
\lambda _{i}\simeq \lim_{T\rightarrow \infty }\frac{1}{T}\sum_{t=1}^{T}\ln \left|
(1-2D)r(i)(1-2q(i,t))\right|.
\end{equation}
It is clear that the effect of diffusion is to stabilize the system. This
may be interesting economically.

Consider single species living in patches $i=1,2,...,n$ with small number
diffusing between the neighboring patches, then the corresponding CML is
\begin{equation}
q_{i}^{t+1}=r_{i}q_{i}^{t}\left( 1-q_{i}^{t}\right)
+D\left(q_{i-1}^{t}+q_{i+1}^{t}-2q_{i}^{t}\right)
\end{equation}
The nonzero steady states are given by Eq. (4), and Lyapunov exponents are given
by Eq. (6). Again diffusion has a stabilizing effect. This confirms that, in
general, spatial diversity stabilizes the biological system.

A more realistic formulation for CML (Eqs. (3), (7)) which takes into account the
fact that interactions have finite propagation speed is \cite {za}
\[
q_{i}^{t+1}=r_{i}q_{i}^{t}\left( 1-q_{i}^{t}\right)
+D\left(q_{i-1}^{t-1}+q_{i+1}^{t-1}-2q_{i}^{t}\right) 
\]
The steady states are not affected by the delay.

Appling the above analysis to the 2-dimensional Lotka-Volterra
predator $(y)$, prey $(x)$ CML \cite {so}
\begin{equation}
\begin{array}{c}
x_{t+1}(i,j) =\mu x_{t}(i,j) \left[1-x_{t}(i,j)-y_{t}(i,j)\right]+ \\

D_{1}\left[x_{t}(i+1,j)+x_{t}(i-1,j)+x_{t}(i,j+1)+x_{t}(i,j-1)-4x_{t}(i,j)\right],\\
 y_{t+1}(i,j) =\beta x_{t}(i,j)y_{t}(i,j)+\\

D_{2}\left[y_{t}(i+1,j)+y_{t}(i-1,j)+y_{t}(i,j+1)+y_{t}(i,j-1)-4y_{t}(i,j)\right],
\end{array}
\end{equation}
where $\mu,\;\beta,\;D_1$ and $D_1$ are some positive constants. The coexistence
steady state ($x\neq 0,\;y\neq 0$) are
\[
x(i,j)=\frac{1}{\beta} ,\;y(i,j)=1-\frac{1}{\beta} -\frac{1}{\mu}.
\]
The largest local Lyapunov exponent is
\begin{equation}
\begin{array}{c}
\lambda (i,j) =\lim_{T\rightarrow \infty }\frac{1}{T}\sum_{t=0}^{T}\ln \left|
f_{t}(i,j)\right|, \\
f_{t}(i,j) =\frac{1}{2}\left[U_{t}(i,j)+\sqrt {U_{t}^{2}(i,j)-4\beta \mu
x_{t}(i,j)y_{t}(i,j)}\right],\\
U_{t}(i,j)=\mu \left( 1-2x_{t}(i,j) \right)+\beta x_{t}(i,j)-4 \left( D_{1}+D_{2}
\right).
\end{array}
\end{equation}\\
{\bf Proposition 1}: If $1\leq r(i)\leq 4\;\forall i=1,2,...n$, then the interval
$[0,1]$ is a local trapping region for the systems (3) and (7).\\
\\
{\bf Proof. }If $\left| q_{t}^{i}\right| \leq 1$, then
\[
\left| q_{t+1}^{i}\right| \leq (1-2D)\left| \Pi _{t}^{i}\right| +D\left|
\Pi _{t}^{i+1}+\Pi _{t}^{i-1}\right| \leq
\frac{r_{i}(1-2D)}{4}+\frac{D(r_{i-1}+r_{i+1})}{4}\leq 1.
\]\\

Also using the results of \cite {af}, then it is direct to prove the following
proposition.\\
\\
{\bf Proposition 2}: If $D$ is sufficient small and $r(i)>4,\;\forall i=1,2,...n$, then the
symbolic dynamics of the systems (3) and (7) is approximated by the direct product of
the symbolic dynamic of the local subsystems.\\

An interesting application of CML to a realistic ecological problem is the
problem of migrating Salmon in a river \cite {pet}. Typically the predators occupy
fixed positions and their number is approximately constant, hence the system can be
modelled by the following equations
\begin{equation}
x_{t+1}^{i}=-\frac{\gamma y^{i}x_{t}^{i}}{1+x_{t}^{i}}+x_{t}^{i-1},
\end{equation}
where $x_{t}^{i}$ is the number of salmon at site $i$ and time $t$, $y^{i}$ is
the number of predator at site $i$ (assumed to be time independent), and
$\gamma$ is a positive constant. The
number of predated salmon in the $i$-th patch is $\gamma y^{i}/(1+x_{t}^{i})$,
hence it decreases as the patch size $(x_{t}^{i})$ increases. This agrees
with observations. The steady state solution of (10) is
\begin{equation}
x^{i}=\frac{1}{2^{i}}\prod_{k=1}^{i}\left[ \sqrt{4x^{k-1}+(1+\gamma
y^{k})^{2}}-(1+\gamma y^{k})\right].
\end{equation}
To study the stability of Eq. (11), one imposes a small perturbation $
x_{t}^{i}=x^{i}+\epsilon _{t}^{i}$. Then it is straightforward to obtain that $
\epsilon _{t}^{i}=\sum_{k=1}^{i} \left[-\gamma y^{k}/(1+x^{k})\right]^{t}$. Since,
typically, $x^{k}\gg y^{k}$, then we have:\\
\\
{\bf Proposition 3}: For all finite values of $n$, the steady states (11) are
stable.\\

Now CML for an epidemic model is constructed. Consider a certain population,
according to the health of each individual, the population is classified to
susceptible (S), infected (I) or recovered (removed) (R). Assuming that only
infectives (and infected) diffuse, then a CML for the susceptible-infected-removed-susceptible
model \cite {ed} (SIRS) can be written as
\begin{equation}
\begin{array}{c}
s_{t+1}^{j}=s_{t}^{j}-\beta s_{t}^{j}I_{t}^{j}+\alpha
R_{t}^{j},\;\;\;R_{t+1}^{j}=R_{t}^{j}+\gamma I_{t}^{j}-\alpha R_{t}^{j},\\
I_{t+1}^{j}=I_{t}^{j}+\beta s_{t}^{j}I_{t}^{j}-\gamma
I_{t}^{j}+D(I_{t}^{j+1}+I_{t}^{j-1}-2I_{t}^{j}),
\end{array}
\end{equation}
where $\alpha,\;\beta$ and $\gamma$ are positive constants.

For the realistic case of rabies, the CML model can be written in the following form:
\begin{equation}
s_{t+1}^{j}=s_{t}^{j}-\beta s_{t}^{j}I_{t}^{j},\;I_{t+1}^{j}=I_{t}^{j}+\beta
s_{t}^{j}I_{t}^{j}-\gamma I_{t}^{j}+D(I_{t}^{j+1}+I_{t}^{j-1}-2I_{t}^{j}).
\end{equation}
The largest local Lyapunov exponent is given by
\[
\lambda _{i}=\lim_{T\rightarrow \infty }\frac{1}{T}\sum_{t=1}^{T}\ln \left|
\frac{a+\sqrt{a^{2}-4b}}{2}\right|, 
\]
where
\[
a=2-\gamma -2D-\beta I_{t}^{j}+\beta s_{t}^{j},\;b=1-2D-\gamma -\beta
I_{t}^{j}(1-2D-\gamma )+\beta s_{t}^{j}. 
\]

Now the CML of schistosomiasis is discussed. Schistosomiasis is one of the major
communicable diseases of public health. It has socio-economic importance in many
developing countries. Moreover, it has been
observed that infection is concentrated in certain regions (centres), hence
the dynamics of the disease depends on the interaction between infected
persons at these regions. This can be represented as CML. We begin by
studying the dynamics at one centre, then later introduce the coupling
between them. Field observations have indicated that there are two periods
(May-June and October-November), where infection of humans by cercaria is highly
probable. This makes difference equations more suitable to model
schistosomiasis with one time step approximately equal six months. The basic
assumptions are:\\
\\
(i) Human population grows exponentially.\\
\\
(ii) Total snail population is approximately constant.\\
\\
(iii) Infected snails do not reproduce.\\
\\
Thus the local equations can be written as
\begin{equation}
H_{t+1}=\alpha H_{t}-\beta P_{t},\;\;P_{t+1}=\gamma
P_{t}(1-\frac{P_{t}}{d})+\lambda C_{t}H_{t},\;\;C_{t+1}=\mu P_{t}-\nu C_{t},
\end{equation}
where $H_{t}$ is the human population at time $t$, $P_{t}$ is the parasite
(adult worm) population at time $t$, $C_{t}$ is the cercaria population at
time $t$, and $\alpha,\;\beta,\;\gamma,\;\lambda,\;\mu,\;\nu,\;d$ are positive
constants. To keep in touch with observations, we will not scale away any of
them. The steady states of this system are
\begin{equation}
\begin{array}{c}
H=\frac{\beta P}{\alpha -1},\;C=\frac{\mu P}{\nu +1},\\
P=0\;{\rm or}\;P=\frac{1/\gamma -1}{\lambda \mu \beta /[\gamma (\nu +1)(\alpha
-1)]-1/d},
\end{array}
\end{equation}
where $\alpha >1,\;0<\gamma <1$. These conditions agree with observations,
since populations in developing countries are increasing. The condition $0<\gamma <1$
means that parasite population increases only through infections by new cercaria.

The stability analysis depends on the eigenvalues of the Jacobian matrix $A$
\[
A=\left[
\begin{array}{ccc}
\alpha & -\beta & 0 \\ 
\lambda C & \gamma (1-\frac{2p}{d}) & \lambda H \\
0 & \mu & -\nu
\end{array}
\right].
\]
The zero solution is unstable, while the nonzero solution is stable if the
following conditions are satisfied:
\begin{equation}
\begin{array}{c}
P>\frac{(\nu +1)(\alpha -1)^{2}}{\beta \lambda \mu},\\
\gamma (1+\alpha )(1-\frac{2P}{d})+\frac{\beta \lambda \mu }{(\nu +1)(\alpha -1)}>0,\\
\left| \alpha \gamma \nu (1-\frac{2P}{d}+\frac{\alpha \beta \lambda \mu P}{\alpha
-1}- \frac{\nu \beta \lambda \mu P}{\nu +1}\right| <1.
\end{array}
\end{equation}
The first inequality is the familiar threshold condition.

There are two ways to resist schistosomiasis namely treatment and snail
fighting. Both decrease the parameter $\lambda$ in Eq. (14). If the treatment
is perfect, then $\lambda \rightarrow 0,\;P=C=0,\;\alpha \rightarrow 1$ and
$H={\rm constant}$. However it is known that no treatment is perfect, hence the
maximum allowed value for $\lambda$ is
\[
\lambda \leq \frac{\gamma (\nu +1)(\alpha -1)}{\beta \mu d}.
\]

Allowing the infected persons to diffuse, one gets the following CML
\begin{equation}
\begin{array}{c}
H_{t+1}^{i}=\alpha H_{t}^{i}-\beta P_{t}^{i}+D(H_{t}^{i-1}+H_{t}^{i+1}-2H_{t}^{i}),\\
P_{t+1}^{i}=\gamma P_{t}^{i}(1-\frac {P_{t}^{i}}{d})+\lambda C_{t}^{i}H_{t}^{i},\\
C_{t+1}^{i}=\mu P_{t}^{i}-\nu C_{t}^{i}.
\end{array}
\end{equation}
In the limit $D\ll 1$, the steady states are given by Eq. (15). From the above
discussions, we conclude the following:\\
\\
(i) Treatment of most infected persons should be done simultaneously to
prevent reinfection through the diffusion between sites. This
agrees with the idea of synchronization mentioned before.\\
\\
(ii) Snail fighting should be done as efficient as possible, since they are
crucial in the disease spread ($\gamma <1)$.\\
\\
(iii) More than one drug (with independent mode of action) should be used
alternatively to avoid drug resistance.\\

An alternative model to the ones used so far to study spatial effects in
epidemics is given by \cite {ll}
\[
\frac {ds_{i}}{dt}=\mu-\mu s_{i}-\sum_{j=1}^{n} \beta_{ij} I_{j} s_{i},
\]
\[
\frac {dI_{i}}{dt}=-(\mu+\gamma) I_{i} + \sum_{j=1}^{n} \beta_{ij} I_{j} s_{i},
\]
where $s_{i} (I_{i})$ is the fraction of susceptible (infected and infective)
individuals in the $i$-th patch, $i=1,2,...,n$, and $\gamma,\;\mu$ are constants,
$\beta_{ij}$ are a measure that an infective from the $j$-th patch will infect a
susceptible from the $i$-th patch. The total population in each patch is assumed
to be constant. It is realistic to assume that
\[
\beta_{ii}\gg \beta_{ij},\; i\ne j.
\]
Expanding $s_{i}=s_{i}^{0}+s_{i}^{1},\; I_{i}=I_{i}^{0}+I_{i}^{1}$ and
linearizing in $\beta_{ij},\; i\ne j$, $s_{i}^{1}$ and $I_{i}^{1}$, one gets the
following equilibrium solutions
\[
s_{i}^{0}=\frac {\mu+\gamma}{\beta_{ii}} < 1,\;\; I_{i}^{0}=\mu \frac
{\beta_{ii}-\mu-\gamma}{\beta_{ii}(\mu+\gamma)},
\]
\[
s_{i}^{1}=-s_{i}^{0}\frac {\sum_{j\ne
i}\beta_{ij}I_{j}^{0}}{\mu+\beta_{ii}I_{i}^{0}-\mu s_{i}^{0}\beta_{ii}/(\mu+\gamma)},
\;\; I_{i}^{1}=-\mu \frac{s_{i}^{1}}{\mu+\gamma}
\]
It is clear that if the patches are different (i.e. $\beta_{ii}$ changes
significantly with $i$), then the equilibrium solutions are asynchronous. Using
the assumption that $\beta_{ii}\gg \beta_{ij},\; i\ne j$, the approximated
solution becomes
\[
s_{i}=1,\; I_{i}=0\; \forall i=1,2,...,n.
\]
This solution is asymptotically stable if the following determinants are
positive
\[
\Delta_{2}=
\left |
\begin{array}{cc}
a_1&1\\
a_3 & a_2
\end{array}
\right |,\;
\Delta_{3}=
\left |
\begin{array}{ccc}
a_1&1 &0\\
a_3 & a_2 & a_1\\
a_5 & a_4 & a_3
\end{array}
\right |,\;\dots
\Delta_{n}=
\left |
\begin{array}{ccccc}
a_1&1 &0 & \dots& 0\\
a_3 & a_2 & a_1 & \dots& 0\\
a_5 & a_4 & a_3 & \dots& 0\\
\vdots&\vdots&\vdots&\dots&\vdots\\
0&0&0&\dots&a_n
\end{array}
\right |,
\]
where
\[
\begin{array}{c}
a_p=\sum_{i_1\ne i_2\ne i_3\ne \dots \ne i_p}(-1)^p \alpha_{i_1} \alpha_{i_2}
\dots \alpha_{i_p-2} (\alpha_{i_1}\alpha_{i_1}-\beta_{i_{p-1} i_p}
\beta_{i_{p} i_{p-1}}),\\
\alpha_{i}=\beta_{ii}-\mu-\gamma.
\end{array}
\]
In this work, terms were kept to the second order in $\beta_{ij},\; i \ne j$.

\section {Chaos control and chaos synchronization in CML}
In the previous sections, it has been shown that chaos exists in some
realistic CML. In some cases chaos is not an acceptable phenomena, hence
chaos control is required. Chaos control in ordinary systems has been
discussed in several references e.g. \cite {ot,gu,ah}. To control chaos in CML, we
propose to increase the 
coupling constant $D$. Using numerical simulations we found that in a logistic
CML (Eq. (3)) with $n=6$, such that only one of them with $r(i)=2$ (the other five oscillators
had $r(j)=4$) was sufficient to control the chaos of the whole CML provided
that $0.5\geq D \geq 0.493$.

In some cases it is not required to control chaos, but to synchronize it \cite
{pec,co}. In the work of Codreanu and Savici \cite {co}, they synchronized chaos of
two identical logistic oscillators
using a third one. Their work has a drawback that there is no bound on the
used control. In many cases this is not realistic. Here their work is
generalized to CML and bounded control is used to synchronize chaos.
Consider the logistic CML (Eq. (3)), with $r(i)=3.7+0.3*{\rm rnd}$, where rnd is a
uniformly distributed random number between zero and one. Thus all
oscillators are chaotic but different. The symmetric case can be studied
similarly. Assume that it is required to synchronize the CML to the a certain
oscillator (without loss of generality call it the $l$-th oscillator). Then to
each oscillator $i\neq l$ apply the following control $u_{i}^{t}$
\begin{equation}
u_{i}^{t+1}=
\left \{
\begin{array}{cc}
u & {\rm if}\; \left| u\right| <u_{\rm max},\\
u_{\rm max} & {\rm if}\; u>u_{\rm max},\\
-u_{\rm max} & {\rm if}\; u<-u_{\rm max},
\end{array}
\right.
\end{equation}
where
\[
u=-r(i)q_{i}^{t}+r(i)(q_{i}^{t})^{2}+r(l)q_{l
}^{t}-r(l)(q_{l}^{t})^{2}+\frac{1}{2}(q_{i}^{t}-q_{l
}^{t}),
\]
and $u_{\rm max}$ is the maximum allowed control. We found numerically that if $n=10$, then $u_{\rm max}=0.09$ is sufficient to
synchronize the CML. Furthermore the coupling constant $D$ was irrelevant to
the value of $u_{\rm max}$, which is contrary to the case of chaos control.\\
\\
{\bf Proposition 4}: There is bounded $u_{\rm max}$, such that the control (18) will
synchronize the CML (3).\\
\\
{\bf Proof. } Since $q_{i}^{t}$ in Eq. (3) is bounded, then $\exists \; q_{\rm max}
<\infty$, such that $q_{i}^{t}\leq q_{\rm max}$. Choosing $u_{\rm max}=q_{\rm max}$, then
\[
u_{i}^{t+1}=-r(i)q_{i}^{t}+r(i)(q_{i}^{t})^{2}+r(l)q_{l
}^{t}-r(l)(q_{l}^{t})^{2}+\frac{1}{2}(q_{i}^{t}-q_{l
}^{t})\;\;{\rm if}\;\left| u_{i}^{t+1}\right| <u_{\rm max}.
\]
Thus the controlled deviation
\[
\Delta _{i}^{t+1}=q_{i}^{t}-q_{l}^{t}+u_{i}^{t},
\]
satisfies
\[
\Delta_{i}^{t+1}=\frac{1}{2}\Delta_{i}^{t}\;\;\mbox{i.e.}\;\; \Delta
_{i}^{t}\rightarrow 0.
\]\\

In practice, simulations have shown that the required $u_{\rm max} \ll q_{\rm max}$.
The previous study is applicable to any CML.

\section{Persistence in some spatially inhomogeneous systems}
Typically a 1-dimension CML is given by
\begin{equation}
\theta _{j}^{t+1}=(1-D)\theta _{j}^{t}+\frac{D}{2} [\theta _{j+1}^{t}+\theta
_{j-1}^{t}]+f(\theta _{j}^{t}),
\end{equation}
or
\begin{equation}
\theta _{j}^{t+1}=(1-D)f(\theta _{j}^{t}\;)\;+\frac{D}{2} [f(\theta
_{j-1}^{t})+f(\theta _{j+1}^{t})],
\end{equation}
where $t=1,2,\dots,\; j=1,2,\dots ,n$.\\
\\
{\bf Definition 1}: A dynamical system is persistent if $\forall \underline{x}(0)\in
{\rm int}(S)$, then $\lim_{t\rightarrow \infty }\inf x_{i}(t)>0 \; \forall i=1,2,\dots,n$.
A CML is persistent if all its components are persistent.\\
\\
An open question is what is the effect of diffusion on persistence?\\
\\
{\bf Proposition 5}: (i) If $\exists \epsilon >0$ such that $\forall \theta \in
\lbrack 0,\epsilon ]$, then $f^{\prime }(\theta )\ge 0$, and $(1-D+f^{\prime
}(\theta ))\geq 0$, then the CML (19) is persistent.\\
\\
(ii) If $f(\theta )\ge 0,\;\forall \theta \geq 0$ and $\exists \epsilon >0$, such that
$\theta \in \lbrack 0,\epsilon ]\Rightarrow f^{\prime }(\theta )\ge \eta >0$, and
$\eta (1-D)\ge 1$, then the CML (20) is persistent.\\
\\
{\bf Proof. } (i) $\forall \theta _{j}^{t}\in (0,\epsilon ]$, using mean value theorem,
then $\exists \zeta _{j}^{t}\in \lbrack 0,\theta _{j}^{t}]$, such that
\[
\theta _{j}^{t+1}=(1-D)\theta _{j}^{t}+\frac{D}{2} [\theta _{j+1}^{t}+\theta
_{j-1}^{t}]+f^{\prime }(\zeta _{j}^{t})\theta _{j}^{t}\ge (1-D)\theta
_{j}^{t}+\frac{D}{2} [\theta _{j+1}^{t}+\theta _{j-1}^{t}].
\]
Hence
\[
\theta _{j}^{t+1}\ge \phi _{j}^{t+1}\equiv (1-D)\theta
_{j}^{t}+\frac{D}{2}[\theta _{j+1}^{t}+\theta _{j-1}^{t}].
\]
The right hand side is a circulant matrix whose eigenvalues are known, and its
maximum eigenvalue is unity. Thus
\[
\phi _{j}^{t+1}=\phi _{j}^{0}\equiv \theta
_{j}^{0}\Rightarrow \theta _{j}^{t+1}\geq \theta _{j}^{0}>0.
\]\\
(ii) For CML (20), the conditions of the proposition imply $\theta _{j}^{t+1}\ge
(1-D)f(\theta _{j}^{t})$. If $\theta _{j}^{t}\in (0,\epsilon ]$, then using
mean value theorem one gets (so long as $\theta _{j}^{s}\in (0,\epsilon
]\;\forall s\leq t$, otherwise $\theta _{j}^{s}>\epsilon$ which implies
persistence), i.e
\[
\theta _{j}^{t+1}\geq (1-D)\eta \theta _{j}^{t}\geq \lbrack (1-D)\eta
]^{t}\theta _{j}^{0} >\theta _{j}^{0}>0.
\]\\

To study persistence in partial differential equations, we use the following
results \cite {cant3} with slight modifications:\\
\\
{\bf Theorem 1}: (i) Suppose that $f(\overrightarrow{r},u)$ is Lipschitz in $
\overrightarrow{r}$ and continuously differentiable in $u$ with
\[
\frac{\partial f}{\partial u} \leq 0\;{\rm for}\;u\geq 0,\;f(\overrightarrow{r},u)\leq
0\;{\rm if}\;u\geq l,
\]
for some constant $l$, and $f(\overrightarrow{r},0)>0$ at some point in the
domain $\Omega$. Then the following problem
\begin{equation}
\frac{\partial u}{\partial t}=D\nabla ^{2}u+uf(\overrightarrow{r}
,u)\;\;\;{\rm in}\;\Omega x(0,\infty ),
\end{equation}
with Dirichlet or Neumann boundary conditions has a unique positive steady
state $u^{ss}$ which is a global attractor for nontrivial non-negative
solutions (hence the system (21) is persistent), if the following problem has
a positive eigenvalue $\sigma $
\begin{equation}
\sigma u=D\nabla ^{2}u+uf(\overrightarrow{r},0)\;\;\;{\rm in}\;\Omega,
\end{equation}
with the same boundary conditions as (21).\\
\\
(ii) Suppose that $f_{1}(\overrightarrow{r},u_{1},u_{2})$ and $f_{2}(
\overrightarrow{r},u_{1},u_{2})$ are $C^{2}$, bounded that $f_{1}(
\overrightarrow{r},u_{1},0)$ satisfies the conditions of part (i), with
positive $\sigma $, and $f_{2}(\overrightarrow{r},0,u_{2})\leq 0$ for $
u_{2}\geq 0$. Let $\sigma _{0}$ be the largest eigenvalue of the system, then
\begin{equation}
\sigma _{0}u_{1}=D_{1}\nabla ^{2}u_{1}+u_{1}f_{1}(\overrightarrow{r}
,0,0)\;\;\;{\rm in}\;\Omega,
\end{equation}
with Dirichlet or Neumann boundary conditions. Also let $\sigma _{2}$ be the
largest eigenvalue of the system
\begin{equation}
\sigma _{2}u_{2}=D_{2}\nabla ^{2}u_{2}\;+u_{2}f_{2}(\overrightarrow{r}
,u_{1}^{ss},0)\;\;\;{\rm in}\;\Omega,
\end{equation}
where $u_{1}^{ss}$ is given by $f_{1}(\overrightarrow{r},u_{1}^{ss},0)=0$,
then the following system is persistent if both $\sigma _{0}$ and $\sigma
_{2}$ are positive
\begin{equation}
\begin{array}{c}
\frac{\partial u_{1}}{\partial t}=D_{1}\nabla ^{2}u_{1}+u_{1}f_{1}(
\overrightarrow{r},u_{1},u_{2}),\\
\frac{\partial u_{2}}{\partial t}=D_{2}\nabla ^{2}u_{2}+u_{2}f_{2}(
\overrightarrow{r},u_{1},u_{2}),
\end{array}
\;\;\;{\rm in}\;\Omega x(0,\infty ).
\end{equation}\\

Now theorem (1) is applied to some 2-dimensional systems. The domain $\Omega$ is
chosen to be the rectangle $x\in \lbrack 0,L_{1}],\;y\in \lbrack
0,L_{2}]$. An example for two species problems is host-parasitoid, where
$u_{1}\; (u_{2})$
is host (parasitoid), respectively and $f_{1}$ and $f_{2}$ in Eq. (25) are given
by
\begin{equation}
f_{1}=b\frac{1-au_{1}}{1+u_{2}}-1,\;\;f_{2}=\frac{u_{1}}{1+u_{2}}-1,
\end{equation}
and persistence conditions become
\begin{equation}
\begin{array}{c}
\frac{b-1}{ab}>1+D_{2}(\frac{\pi ^{2}}{L_{1}^{2}}+\frac{\pi ^{2}}{L_{2}^{2}}),\\
b-1>D_{1}(\frac{\pi ^{2}}{L_{1}^{2}}+\frac{\pi ^{2}}{L_{2}^{2}}).
\end{array}
\end{equation}

Another example is the predator-prey or epidemic model where $u_{1}$ ($u_{2}$)
is prey (predator) or susceptible (infected and infectious) and $
f_{1}$ and $f_{2}$ in Eq. (25) are
\begin{equation}
f_{1}=1-u_{1}-u_{2},\;\;\;f_{2}=u_{1}-\gamma,
\end{equation}
and persistence conditions are
\begin{equation}
1>D_{1}(\frac{\pi ^{2}}{L_{1}^{2}}+\frac{\pi ^{2}}{L_{2}^{2}})\; {\rm and}\; 1>\gamma
+D_{2}(\frac{\pi ^{2}}{L_{1}^{2}}+\frac{\pi ^{2}}{L_{2}^{2}}).
\end{equation}

For the case of rabies one sets $D_{1}=0$, since only infected foxes spread,
while susceptible ones do not. This indicates that a method to fight
infection spread is to put susceptibles in small groups (to decrease the
domain).

Now the Krill-Whale model \cite {be} will be studied
including diffusion. The functions $f_{1}$ and $f_{2}$ in Eq. (25) are
\[
f_{1}=ru_{1}(1-u_{1})-u_{1}u_{2},\;\;\;f_{2}=su_{2}\frac{1-u_{2}}{u_{1}},
\]
where $u_{1}$ ($u_{2}$) represents krill (whale), respectively and $r$, $s$ are
positive constants. The uniform coexistence steady states are $
u_{1}=1-1/r$, $u_{2}=1$, $r>1$. Its stability requires $s<1-1/r$, 
$1<r<2$. Sufficient conditions for persistence can
be derived (using theorem (1)) in the following form
\begin{equation}
r>D_{1}(\frac{\pi ^{2}}{L_{1}^{2}}+\frac{\pi
^{2}}{L_{2}^{2}})+1,\;\;\;s>D_{2}(\frac{\pi ^{2}}{L_{1}^{2}}+\frac {\pi ^{2}}{L_{2}^{2}})
\end{equation}
Furthermore using Dulac-Bendixon lemma, one can see that this system has
no cycles.

There is a weakness in the definition of persistence in that it does not
guarantee that the population is larger than a prescribed value. To find the
minimum of the population is part of the practical persistence \cite {cao,cant6}.
Studying this concept for the krill-whale model, one can derive the following limits
for the populations
\[
1\geq u_{1}\geq 1-\frac{1}{r},\;\;\;u_{2}\simeq 1.
\]
These values agreed with the numerical simulations. 

One may argue that advection should be included in the above diffusion
model. So an interesting question arises as to what is the effect of
advection on the persistence conditions of theorem (1)? Consider the following 
diffusion-advection system
\begin{equation}
\frac{\partial u}{\partial t}=D\nabla ^{2}u+uf(\overrightarrow{r},u)-v \frac{\partial
u}{\partial x}\;\;\;{\rm in}\;\Omega x(0,\infty ).
\end{equation}
If the system (31) admits a solution of the following form
\begin{equation}
u(x,t)=\sum_{n=1}e^{\lambda _{n}t}(a_{n}\sin \frac{n\pi x}{L}+b_{n}\cos \frac{n\pi
x}{L}),
\end{equation}
and linearizing around $u=0$, one gets that the advection term in Eq. (31) affects
only the imaginary part of the eigenvalue $\sigma$ in Eq. (22). Hence we
conclude the following proposition:\\
\\
{\bf Proposition 6}: If the system (31) admits a solution of the form (32), then
the persistence conditions of theorem (1) are valid.

\section{A method to study differential equations with periodic
coefficients}
We begin by proposing the following correspondence between differential
equations with periodic coefficients and some difference equations. Consider
the following system
\begin{equation}
\frac{dx_{i}}{dt}=\sum_{j}p_{ij}(t)x_{j},\;\;\;p_{ij}(t)=p_{ij}(t+T).
\end{equation}
Approximating the left hand side using
\[
\frac{dx_{i}}{dt}\simeq \frac{x_{i}(t+T)-x_{i}(t)}{T},
\]
thus one gets
\begin{equation}
x_{i}(t+T)=x_{i}(t)+T\sum_{j}p_{ij}(t)x_{j},\;\;\;t=0,T,2T,\dots.
\end{equation}
Since the discrete system (34) has time step $T$, then $p_{ij}(t)$ will be
constant in its evolution, thus the system (34) is autonomous.

Recalling that the stability conditions for discrete systems are more
stringent than the corresponding continuous one, then stability conditions
for the system (34) will be sufficient conditions for those of (33). Thus we have\\
\\
{\bf Proposition 7}: If all the eigenvalues $\lambda$ of the matrix $[\delta
_{ij}+Tp_{ij}(t)]$ satisfy $\left| \lambda \right| \leq 1\;\forall t\in
\lbrack 0,T]$, then the periodic solution of Eq. (33) is stable.

As an example consider the following equation
\begin{equation}
\frac{d^{2}x}{dt^{2}}+p(t)x=0,\;\;\;p(t+T)=p(t).
\end{equation}
Approximating it by
\begin{equation}
x(t+T)+x(t-T)+x(t)(-2+T^{2}p(t))=0,\;\;\;t=T,\;2T,\;3T,\;\dots.
\end{equation}
Notice that in Eq. (36), $p(t)$ is constant since $t$ increases by $T$, and
$p(t)=p(t+T)$. Thus one gets that Eq. (36) is stable if
\begin{equation}
0<T^{2}p(t)<4\;\forall t\in \lbrack
0,T],
\end{equation}
which yields the known Lyapunov theorem for the sufficient
conditions for the stability of the periodic motion of Eq. (36)
\begin{equation}
p(t)>0,\;0<T\int_{0}^{T}p(s)ds<4.
\end{equation}

Applying these results to Lame's equation
\[
\frac{d^{2}x}{dt^{2}}-(h+2k^{2}{\rm sn}^{2}t)x=0,
\]
one gets the following sufficient conditions for the stability of its periodic
motion
\begin{equation}
\begin{array}{c}
0\leq -4K^{2}(h+2k^{2}{\rm sn}^{2}t)\leq 4,\;\forall t\in \lbrack
0,2K],\\
K=\int_{0}^{\pi /2}\frac{d\phi}{\sqrt{1-k^{2}\sin ^{2}\phi }}.
\end{array}
\end{equation}

Applying them to Hill's equation
\[
\frac{d^{2}x}{dt^{2}}+(a-2b\cos 2\pi t)x=0,
\]
one gets that the sufficient conditions for the stability of its periodic motion are
\begin{equation}
0<a-2b\cos 2\pi t<4\;\forall t\in \lbrack 0,1].
\end{equation}

The previous method is also useful for the case of periodically forced Lotka-Volterra
predator prey models of the following form
\begin{equation}
\frac{dp}{dt}=-\alpha p+\beta pv,\;\frac{dv}{dt}=(r+\epsilon \sin 2\pi
t)v(1-\frac{\nu}{K})-\gamma p v.
\end{equation}
Following the above procedure, one gets the following stability
conditions:
\begin{equation}
0<\alpha \frac{r+\epsilon \sin 2\pi t}{\beta K}<4\;\ \forall t\in \lbrack
0,1].
\end{equation}

This method is expected to simplify the study of seasonality.

\section{Conclusions}
In conclusion, ecologists have stated that it is important to include several
factors in some biological and epidemic models e.g. spatial heterogeneity,
synchronization and seasonality. Since a crucial question is whether the
epidemic (or the population) will persist, we studied the persistence of some
spatially heterogenous systems (using CML or partial differential
equations). Synchrony in CML and CA is studied. Since chaos implies
asynchrony, we studied bounded chaos control on CA and CML. A simple method
to study seasonally in forced systems (systems with periodic coefficients) is
introduced.

\section*{Acknowledgements}
We are grateful to R. Cantrell and C. Cosner for sending some copies of
their work to us and for discussions.


\begin{thebibliography}{99}

\bibitem{af} Afraimovich V. and Fernandez B. (2000), Nonlinearity 13, 973.  

\bibitem{ah} Ahmed E., Agiza H.N. and Hassan S.Z. (1999) Chaos Solitons and
Fractals 10, 1179.

\bibitem{be} Beddington A and May R.M. (1982), "The harvesting of interacting
species in a natural ecosystem", Sci. Amer. 62-69.

\bibitem{bo} Boccara N., Goles E., Martinez S. and Picco P. (eds) (1993), "Cellular
automata and cooperative phenomena", Kluwer, New York.

\bibitem{cant3} Cantrell R.S, Cosner C. and Hutson V. (1993), Proc. Roy. Soc.
Edin. 123A, 533.

\bibitem{cant6} Cantrell R.S and Cosner C. (1996), Proc. Roy. Soc. Edin. 126A, 247.

\bibitem{cao} Cao Y. and Gard T.C. (1993), Differential Integ. Equations 6, 883.

\bibitem{co} Cordeanu S. and Savici A. (2001), Chaos Solitons and Fractals 12, 845.

\bibitem{cr} Cross M.C. and Hohenberg P.C. (1993), Rev. Mod. Phys 65, 851.

\bibitem{ea} Earn D.J.D., Rohani P. and Grenfell B.T. (1998), Proc. R. Soc. London B 265, 7.

\bibitem{ed} Edelstein-Keshet L. (1988), "Mathematical Models in Biology", Random House,
New York.

\bibitem{gu} Gumez J. and Matias M.A. (1996) Phys. Rev. E 53, 3059.

\bibitem{ho} Holmgren R. (1996), "An Introduction to Discrete Dynamical
Systems" Springer, Berlin.

\bibitem{ka} Kaneko K.(1993), "Theory and Application of Coupled Map Lattices",
J.Wiley Publisher, New York.

\bibitem{ll} Lloyd A. and May R.M. (1996), J. Theor. Biol. 179, 1.

\bibitem{ot} Ott E., Grebogi C. and Yorke J.A.(1990), Phys. Rev. Lett. 64, 1193.

\bibitem{pec} Pecora C.M. and Carrol T.L. (1990), Phys. Rev. Lett. 64, 821.

\bibitem{pet} Petersen J.H. and De Angelis D.L. (2000), Math. Biosci. 165, 97.

\bibitem{ra} Rai V. and Schaffer W.M. (2001), Chaos Solitons and Fractals 12,
197.

\bibitem{so} Sole V. and Valls J. (1991), Phys. Lett. A 153, 330.

\bibitem{za} Zanette D.H. (2000), Cond-mat/0003174.

\end{thebibliography}
\end{document}